\documentclass[12pt]{article}
\usepackage{amssymb,amsmath,epsfig}

\begin{document}

\title{\bf Cosmological Evolution of Interacting New Holographic
Dark Energy in Non-flat Universe}
\author{M. Sharif \thanks {msharif.math@pu.edu.pk} and Abdul Jawad\thanks {jawadab181@yahoo.com}\\
Department of Mathematics, University of the Punjab,\\
Quaid-e-Azam Campus, Lahore-54590, Pakistan.}

\date{}

\maketitle
\begin{abstract}
We consider the interacting holographic dark energy with new
infrared cutoff (involving Hubble parameter and its derivative) in
non-flat universe. In this context, we obtain the equation of state
parameter which evolutes the universe from vacuum dark energy region
towards quintessence region for particular values of constant
parameters. It is found that this model always remains unstable
against small perturbations. Further, we establish the
correspondence of this model having quintessential behavior with
quintessence, tachyon, K-essence and dilaton scalar field models.
The dynamics of scalar fields and potentials indicate accelerated
expansion of the universe which is consistent with the current
observations. Finally, we discuss the validity of the generalized
second law of thermodynamics in this scenario.
\end{abstract}

\section{Introduction}

Dark energy (DE) is one of the most attractive and active fields in
modern cosmology due to the indications of accelerated expansion of
the universe through type Ia Supernovae \cite{1}. Observational data
like CMBR \cite{2}, large scale structure \cite{3,4}, gravitational
lensing surveys \cite{5} and galaxy redshift surveys \cite{6} also
favor this phenomenon. However, the identity of DE is still
ambiguous and various models have been suggested to know its nature.
Its simplest candidate is the cosmological constant but it suffers
two well-known problems, i.e., "fine tuning problem" and "cosmic
coincidence problem" \cite{7}.

In order to understand the nature of DE phenomenon, various
dynamical DE models have been proposed which can be characterized by
the equation of state (EoS) parameter $\omega$. The holographic dark
energy (HDE) is one of the emergent dynamical DE model proposed in
the context of fundamental principle of quantum gravity, so called
holographic principle \cite{8}. It is derived with the help of
entropy-area relation of thermodynamics of black hole horizons in
general relativity which is also known as the Bekenstein-Hawking
entropy bound, i.e., $S\simeq M^2_{p}L^2$, where $S$ is the maximum
entropy of the system of length $L$ and $M_{p}=(8\pi
G)^{-\frac{1}{2}}$ is the reduced Planck mass. Using this relation,
Cohen et al. \cite{9} argued that the vacuum energy (or the quantum
zero-point energy) of a system with size $L$ should always remain
less than the mass of a black hole with the same size due to the
formation of black hole in quantum field theory. Hsu \cite{10} and
Li \cite{11} formulated this statement in mathematical form as
\begin{equation*}
\rho_{\Lambda}=\frac{3m^{2}}{8\pi GL^{2}},
\end{equation*}
which is known as HDE density, $m$ is constant, $L$ is the infrared
(IR) cutoff and $G$ is the gravitational constant.

The density of HDE describes the connectivity between ultraviolet
(UV) and IR cutoffs which represent the bounds of energy density and
size of the universe, respectively. The HDE model suffers the choice
of IR cutoff problem. Li \cite{11} proved that Hubble as well as
particle horizons are not compatible with the present status of the
universe while the future event horizon is the best candidate for
non-interacting HDE with suitable parameter $m$. It is argued
\cite{12} that HDE with future event horizon plagued with causality
problem (why should we calculate the current value of DE density
with the help of future event horizon of the universe$?$). This
problem motivated people \cite{12}-\cite{14} to modify the IR cutoff
as a function of the Ricci scalar or generalized form of the Ricci
scalar. Further, observational analysis has also been done for these
types of HDE models \cite{15}.

The stability against small perturbations is a well-known procedure
to check the viability of a DE model. For this purpose, the sign of
the square of the speed of sound,
$\upsilon^2_s=\frac{dp_\Lambda}{d\rho_\Lambda}$, plays a key role -
its negativity represents instability and vice versa \cite{7}. It
was shown \cite{17} that Chaplygin and tachyon Chaplygin gases are
stable as $\upsilon^2_s>0$. However, for the holographic \cite{18},
agegraphic \cite{19} and QCD ghost DE \cite{20} models,
$\upsilon^2_s<0$, i.e., they are classically unstable.

The scalar-field dark energy model then can be considered as an
effective description of this holographic theory. The reconstruction
of HDE in terms of scalar fields has been discussed widely. The
scalar fields (which naturally occur in particle physics such as
string theory \cite{21}) are used as a possible candidate of DE. In
this scenario, a large number of models have been proposed including
quintessence, phantom, K-essence, tachyon, ghost condensates and
dilatonic DE \cite{21,22} etc. Granda and Oliveros \cite{23}
formulated the scalar field models for HDE by using new IR cutoff
\cite{13} (called new HDE (NHDE)) in flat FRW universe. Karami and
Fehri \cite{24} generalized this work to the non-flat universe.
Recently, Sheykhi \cite{27} has constructed the quintessence,
tachyon, K-essence and dilatonic DE scalar field models for
interacting HDE with Hubble horizon as an IR cutoff in flat
universe. The generalized second law of thermodynamics (GSLT) was
also discussed for HDE model with different IR cutoff for non-flat
universe \cite{25}.

It is believed that an early inflation era provides a flat universe,
but this consequence is only true if the number of e-foldings is
very large \cite{25a}. The data of first year WMAP analysis favors
the non-flat scenario of the universe \cite{25b}. Also, different
observational data provided the evidence about the contribution of
spatial curvature to the total energy density of the universe
\cite{25c}-\cite{25i}. It was shown that the parameterizations of
the dark energy models admit the non-flat universe by implying
compatible observational data \cite{25j}. Recently, Lu et al.
\cite{25k} have used type Ia supernovae, baryon acoustic
oscillations, CMBR and observational Hubble data and obtained the
value $-0.0013^{+0.0130}_{-0.0040}$ for fractional energy density
due to curvature. It would be interesting to study the universe with
a spatial curvature.

In view of above discussion, we extend the work of Sheykhi \cite{27}
by taking the generalized form of IR cutoff of HDE in non-flat
universe. We also check the validity of the GSLT in this scenario.
The paper is organized as follows: Section \textbf{2} contains the
discussion of the evolution and instability of interacting NHDE in
non-flat universe. Section \textbf{3} is devoted for the
reconstruction of scalar field models of interacting NHDE while
section \textbf{4} investigates the validity of GSLT. The last
section provides the summary of our results.

\section{New Holographic Dark Energy}

In this section, we manipulate the expressions of EoS parameter and
speed of sound for NHDE interacting with dark matter (DM) in
non-flat FRW universe
\begin{equation}\label{1}
ds^{2}=-dt^{2}+a^{2}(t)[\frac{dr^{2}}{1-kr^{2}}
+r^{2}(d\theta^{2}+\sin\theta^{2}d\phi^{2})].
\end{equation}
Here $a(t)$ is the cosmic scale factor which measures the expansion
of the universe and $k=-1,0,1$ represents the spatial curvature
indicating the open, flat and closed universes, respectively. The
corresponding equations of motion are
\begin{eqnarray}\label{2}
H^2+\frac{k}{a^{2}}&=&\frac{1}{3}(\rho_m+\rho_\Lambda),\\\label{3}
\dot{H}+H^2&=&-\frac{1}{6}(\rho_m+\rho_\Lambda+3p_\Lambda),
\end{eqnarray}
here we assume $8\pi G=1$ as well as dust like DM. Also, $\rho_m$
and $\rho_\Lambda$ are DM and DE densities, $p_\Lambda$ is pressure
due to DE, $H$ denotes the Hubble parameter and dot represents
derivative with respect to time. We can rewrite Eq.(\ref{3}) in the
form of fractional energy densities as
\begin{equation}\label{4}
\Omega_{m}+\Omega_{\Lambda}=1+\Omega_{k},
\end{equation}
where
\begin{equation*}\label{5}
\Omega_{m}=\frac{\rho_{m}}{3H^{2}},\quad
\Omega_{\Lambda}=\frac{\rho_{\Lambda}}{3H^{2}},\quad
\Omega_{k}=\frac{k}{a^2H^{2}}.
\end{equation*}
We assume the NHDE density in the following form \cite{13}
\begin{equation}\label{6}
\rho_{\Lambda}=3(\mu H^2+\lambda\dot{H}),
\end{equation}
where $\mu$ and $\lambda$ are positive constants.

The interaction of DM and NHDE leads to the equation of continuity
in the form of two non-conserving equations as
\begin{eqnarray}\label{7}
\dot{\rho}_{m}+3H\rho_{m}&=&\Upsilon,\\\label{8}
\dot{\rho}_{\Lambda}+3H(\rho_{\Lambda}+p_{\Lambda})&=&-\Upsilon.
\end{eqnarray}
We choose the following form of the interaction term $\Upsilon$
\begin{equation}\label{9}
\Upsilon=3d^{2}H\rho_{m},
\end{equation}
here $d^2$ is the interacting constant. Using Eqs.(\ref{7}) and
(\ref{9}), we obtain DM density as follows
\begin{equation}\label{10}
\rho_{m}=\rho_{m_{0}}e^{-3(1-d^2)x},
\end{equation}
where $x=\ln a$ and $\rho_{m_0}$ is an integration constant.
Inserting Eqs.(\ref{6}) and (\ref{10}) in Eq.(\ref{2}), we obtain
the differential equation
\begin{equation}\label{11}
\frac{dE^2}{dx}+\frac{2(\mu-1)}{\lambda}E^2=\frac{2\Omega_{k_0}}{\lambda}e^{-2x}
-\frac{2\Omega_{m_0}}{\lambda}e^{-3(1-d^2)x},
\end{equation}
where $E^2=\frac{H^2}{H^2_{0}},~(0)$ represents the present value of
the parameter and $b$ is an integration constant. This has the
solution
\begin{equation}\label{12}
E^2=\frac{\Omega_{k_0}}{\mu-\lambda-1}e^{-2x}
-\frac{2\Omega_{m_0}}{2(\mu-1)-3(1-d^2)\lambda}e^{-3(1-d^2)x}+b
e^{-\frac{2}{\lambda}{(\mu-1)x}}.
\end{equation}
The initial condition $a_0=1$ yields $x=0,~E=1$, and hence we have
\begin{eqnarray}\label{13}
b=1-\frac{\Omega_{k_{0}}}{\mu-\lambda-1}
+\frac{2\Omega_{m_{0}}}{2(\mu-1)-3(1-d^2)\lambda}.
\end{eqnarray}
\begin{figure} \centering
\epsfig{file=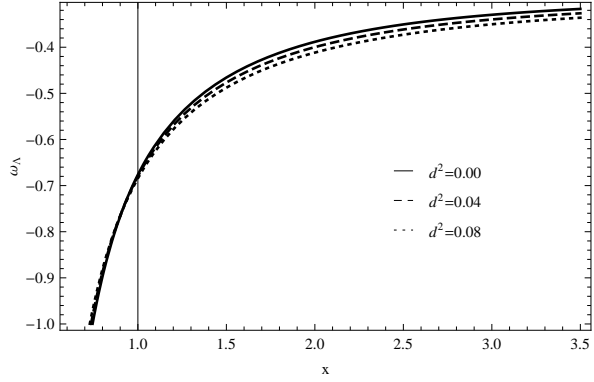,width=.60\linewidth} \caption{Plot of
$\omega_{\Lambda}$ versus $x$ for NHDE.}
\end{figure}

With the help of Eqs.(\ref{6}), (\ref{8}) and (\ref{12}), we obtain
the evolution parameter of interacting NHDE as
\begin{eqnarray}\nonumber
\omega_{\Lambda}&=&\left[\frac{(\lambda-\mu)\Omega_{k_{0}}}{3(\mu-\lambda-1)}e^{-2x}
+\frac{2d^2\Omega_{m_{0}}}{2(\mu-1)-3(1-d^2)\lambda}e^{-3(1-d^2)x}-\frac{b}{3\lambda}\right.\\\nonumber
&\times&\left.(3\lambda-2\mu+2)e^{\frac{-2(\mu-1)x}{\lambda}}\right]\left[\frac{(\mu
-\lambda)\Omega_{k_{0}}}{(\mu-\lambda-1)}e^{-2x}-(2\mu-3(1-d^2)\lambda)\right.\\\label{14}
&\times&\left.(2(\mu-1)-3(1-d^2)\lambda)^{-1}\Omega_{m_{0}}e^{-3(1-d^2)x}
+be^{\frac{-2(\mu-1)x}{\lambda}}\right]^{-1}.
\end{eqnarray}
We plot the EoS parameter $\omega_{\Lambda}$ versus $x$ with respect
to different well-known choices of interacting parameter $d^2$,
i.e., 0, 0.04, 0.08 shown in Figure \textbf{1}. Also, we assume the
NHDE parameters as $\mu=1.198,~\lambda=0.195$ and the current values
of $\Omega_{k_0}=0.01,~\Omega_{\Lambda_0}=0.73$. We see that the EoS
parameter translates the universe from vacuum DE region towards
quintessence region.
\begin{figure} \centering
\epsfig{file=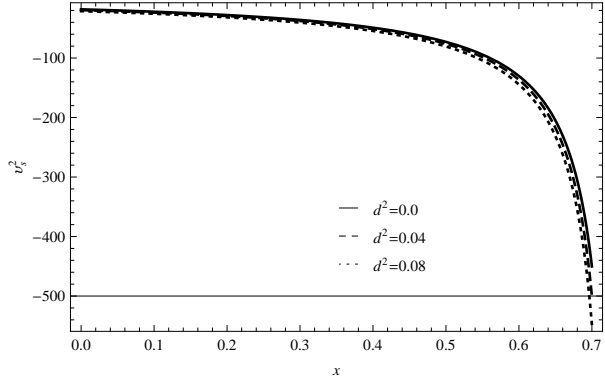,width=.60\linewidth} \caption{Plot of
$\upsilon_{s}^2$ versus $x$ for NHDE.}
\end{figure}

The speed of sound is given by \cite{7}
\begin{equation}\label{15}
\upsilon_{s}^2=\frac{\dot{p}}{\dot{\rho}}=\frac{p'}{\rho'},
\end{equation}
where prime means differentiation with respect to $x$. Using
Eqs.(\ref{6}), (\ref{8}), (\ref{12}) and (\ref{15}), it follows
that
\begin{eqnarray}\nonumber
\upsilon^2_{s}&=&\left[\frac{2(\mu-\lambda)\Omega_{k_{0}}}{3(\mu-\lambda-1)}e^{-2x}
-\frac{6d^2(1-d^2)\Omega_{m_{0}}}{2(\mu-1)-3(1-d^2)\lambda}e^{-3(1
-d^2)x}+2b(\mu-1)\right.\\\nonumber
&\times&\left.\frac{(3\lambda-2\mu+2)}{3\mu^2}e^{\frac{-2(\mu-1)x}{\lambda}}\right]
\left[\frac{-2(\mu-\lambda)\Omega_{k_{0}}}{(\mu-\lambda-1)}e^{-2x}
+(2\mu-3(1-d^2)\lambda)\right.\\\nonumber
&\times&\left.3(1-d^2)(2(\mu-1)-3(1-d^2)\lambda)^{-1}\Omega_{m_{0}}e^{-3(1-d^2)x}
-\frac{2b(1-\mu)}{\lambda}\right.\\\label{16}&\times&\left.
e^{\frac{-2(\mu-1)x}{\lambda}}\right]^{-1},
\end{eqnarray}
which provides the speed of sound for interacting NHDE. This is
shown in Figure \textbf{2} for the same constant parameters as given
above. In this scenario, the speed of sound remains negative with
the increase of interacting parameter. This shows that the NHDE is
unstable just like non-interacting HDE model with future event
horizon \cite{18}.

\section{Reconstruction of New Holographic Scalar Field Models}

Here, we provide the correspondence of the interacting NHDE with
quintessence, tachyon, K-essence and dilaton field models in
non-flat universe.

\subsection{New Holographic Quintessence Model}

The ordinary scalar field $\phi$ is governed by quintessence which
is minimally coupled with gravity. The energy density and pressure
of the quintessence scalar field are defined as \cite{22}
\begin{eqnarray}\label{17}
\rho_{q}=\frac{1}{2}\dot{\phi}^{2}+V(\phi),\quad
p_{q}=\frac{1}{2}\dot{\phi}^{2}-V(\phi),
\end{eqnarray}
where $\dot{\phi}^{2}$ and $V(\phi)$ are termed as kinetic energy
and scalar potential, respectively. The EoS parameter for this model
becomes
\begin{equation*}
\omega_{q}=\frac{\dot{\phi}^{2}-2V(\phi)}{\dot{\phi}^{2}+2V(\phi)}.
\end{equation*}
For the correspondence between NHDE and quintessence scalar field,
we set $\rho_{q}=\rho_{\Lambda}$ and $p_{q}=p_{\Lambda}$.
Consequently, Eq.(\ref{17}) yields
\begin{eqnarray}\nonumber
\dot{\phi}^{2}&=&3H^{2}_{0}\left[\frac{2(\mu-\lambda)\Omega_{k_{0}}}{3(\mu-\lambda-1)}e^{-2x}
-\frac{(2\mu-3(1-d^2)\lambda-2d^2)\Omega_{m_{0}}}{2(\mu-1)-3(1-d^2)\lambda}e^{-3(1
-d^2)x}\right.\\\label{19}
&+&\left.\frac{2b(\mu-1)}{3\lambda}~e^{\frac{-2(\mu-1)x}{\lambda}}\right],\\\nonumber
V(\phi)&=&3H^{2}_{0}\left[\frac{2(\mu-\lambda)\Omega_{k_{0}}}{3(\mu-\lambda-1)}e^{-2x}
-\frac{(2\mu-3(1-d^2)\lambda+2d^2)\Omega_{m_{0}}}{2(2(\mu-1)-3(1-d^2)\lambda)}e^{-3(1
-d^2)x}\right.\\\label{20}
&+&\left.\frac{b(3\lambda-\mu+1)}{3\lambda}e^{\frac{-2(\mu-1)x}{\lambda}}\right].
\end{eqnarray}
The kinetic energy can also be written as
\begin{eqnarray}\nonumber
\phi'(a)&=&\sqrt{3}\left[\frac{2(\mu-\lambda)\Omega_{k_{0}}}{3(\mu-\lambda-1)}e^{-2x}
-\frac{(2\mu-3(1-d^2)\lambda-2d^2)\Omega_{m_{0}}}{2(\mu-1)-3(1-d^2)\lambda}e^{-3(1
-d^2)x}\right.\\\nonumber
&+&\left.\frac{2b(\mu-1)}{3\lambda}e^{\frac{-2(\mu-1)x}{\lambda}}\right]^\frac{1}{2}
\left[{\frac{\Omega_{k_{0}}}{(\mu-\lambda-1)}e^{-2x}
-(2\mu-3(1-d^2)\lambda)}\right.\\\label{21}
&\times&\left.(2(\mu-1)-3(1-d^2)\lambda)^{-1}\Omega_{m_{0}}e^{-3(1-d^2)x}+
be^{\frac{-2(\mu-1)x}{\lambda}}\right]^{-\frac{1}{2}}.
\end{eqnarray}
\begin{figure} \centering
\epsfig{file=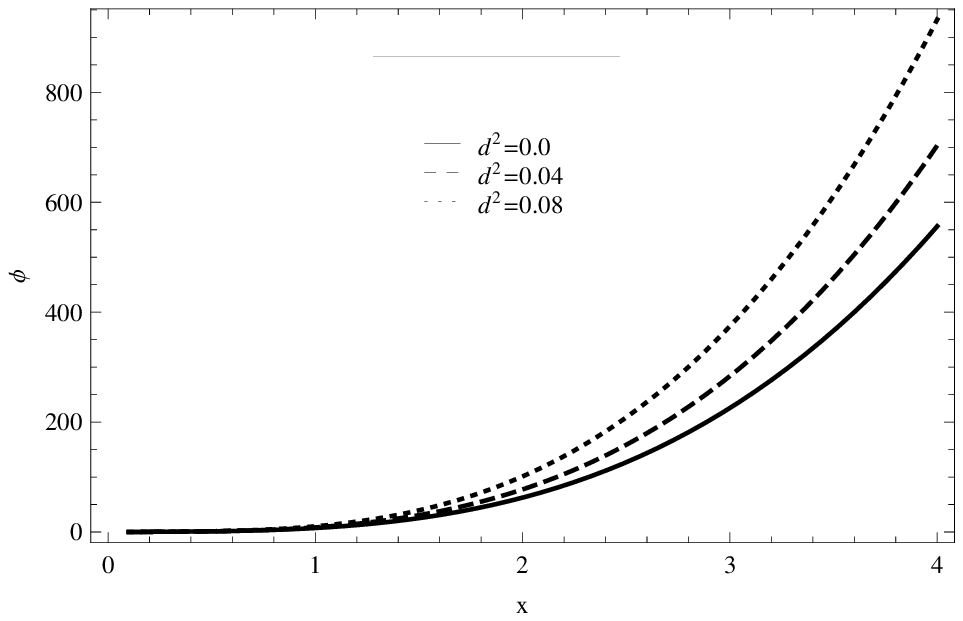,width=.60\linewidth}\caption{Plot of $\phi$
versus $x$ for quintessence model.}
\end{figure}
\begin{figure} \centering
\epsfig{file=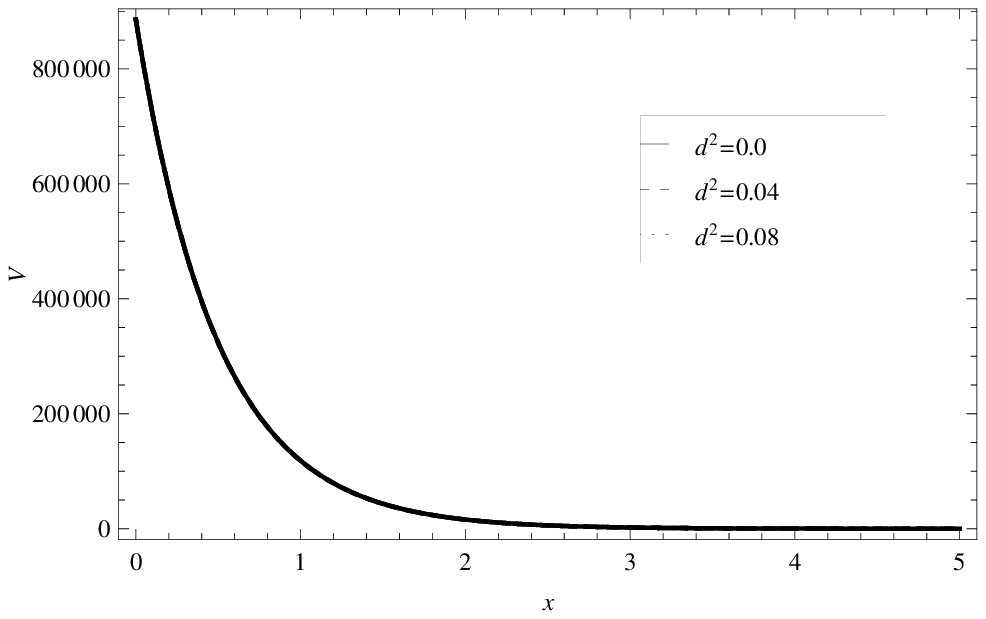,width=.60\linewidth} \caption{Plot of $V$ versus
$x$ for quintessence model.}
\end{figure}
\begin{figure} \centering
\epsfig{file=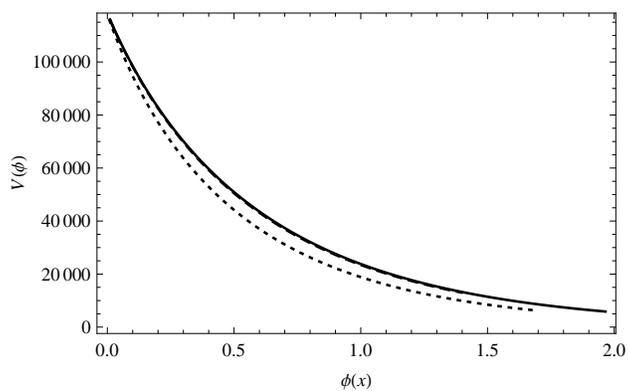,width=.60\linewidth} \caption{Plot of $V$ versus
$\phi$ for quintessence model.}
\end{figure}

This equation cannot be solved analytically to obtain $\phi$ due to
its complicated nature. To get insights, we solve it numerically and
plot $\phi$ against $x$ by choosing initial condition, $\phi(0)=0$,
while keeping the remaining parameters same as in the previous
section. Figure \textbf{3} shows that the scalar field increases
(and hence the kinetic energy $\dot{\phi}^2$ of the potential
decreases) with the passage of time. The potential $V$ versus $x$
and $\phi$ are shown in Figures \textbf{4} and \textbf{5},
respectively, indicating the decreasing behavior. The quintessence
potential in terms of $\phi$ shows large value at the present epoch
which represents accelerated expansion of the universe. This is
consistent with the result shown through phase space analysis
\cite{22} that the exponential potential for the scalar field
contains the attractor solutions describing the accelerated
expansion of the universe. However, in the later time, the universe
remains in the accelerated phase because quintessence potential goes
to positive and non-zero minima but kinetic energy goes to zero.

\subsection{New Holographic Tachyon Model}

The tachyon model, originated from the string theory, has been
suggested to explain DE scenario. It has an interesting feature that
a rolling tachyon interpolates the EoS parameter between $-1$ to
$0$. Also, the tachyon model is the best candidate for inflation at
high energy. Many attempts have been made to formulate reliable
cosmological models with the help of different self-interacting
potentials \cite{28}. However, the effective Lagrangian for this
model is defined as \cite{22}
\begin{equation*}
L=-V(\phi)\sqrt{1+\partial_{\mu}\phi\partial^{\mu}\phi},
\end{equation*}
where $V(\phi)$ represents the tachyon potential. This scalar field
has the following energy and pressure
\begin{eqnarray}\label{22}
\rho_{t}=\frac{V(\phi)}{\sqrt{1-\dot{\phi}^{2}}},\quad
p_{t}=-V(\phi)\sqrt{1-\dot{\phi}^{2}}
\end{eqnarray}
and the EoS parameter is
\begin{equation}\label{24}
\omega_{t}=\dot{\phi}^{2}-1.
\end{equation}
The correspondence between NHDE and tachyon model is obtained for
$\rho_{t}=\rho_{\Lambda}$ and $p_{t}=p_{\Lambda}$ which lead to
\begin{eqnarray}\nonumber
\dot{\phi}^{2}&=&\left[\frac{2(\mu-\lambda)\Omega_{k_{0}}}{3(\mu-\lambda-1)}e^{-2x}
-\frac{(2\mu-3(1-d^2)\lambda-2d^2)\Omega_{m_{0}}}{2(2(\mu-1)-3(1-d^2)\lambda)}e^{-3(1
-d^2)x}\right.\\\nonumber
&+&\left.\frac{2b(\mu-1)}{3\lambda}e^{\frac{-2(\mu-1)x}{\lambda}}\right]
\left[\frac{(\mu-\lambda)\Omega_{k_{0}}}{(\mu-\lambda-1)}e^{-2x}
-(2\mu-3(1-d^2)\lambda)\right.\\\label{25}
&\times&\left.(2(\mu-1)-3(1-d^2)\lambda)^{-1}\Omega_{m_{0}}e^{-3(1-d^2)x}+
be^{\frac{-2(\mu-1)x}{\lambda}}\right]^{-1},\\\nonumber
V(\phi)&=&3H^{2}_{0}\left[\frac{(\mu-\lambda)\Omega_{k_{0}}}{(\mu-\lambda-1)}e^{-2x}
-\frac{(2\mu-3(1-d^2)\lambda)\Omega_{m_{0}}}{2(\mu-1)-3(1-d^2)\lambda}e^{-3(1
-d^2)x}\right.\\\nonumber
&+&\left.be^{\frac{-2(\mu-1)x}{\lambda}}\right]^{\frac{1}{2}}
\left[\frac{(\mu-\lambda)\Omega_{k_{0}}}{3(\mu-\lambda-1)}e^{-2x}
-2d^2~\Omega_{m_{0}}e^{-3(1 -d^2)x}(2(\mu-1)\right.\\\label{26}
&-&\left.3(1-d^2)\lambda)^{-1}
+\frac{b}{3\lambda}(3\lambda-2\mu+2)e^{\frac{-2(\mu-1)x}{\lambda}}\right]^{\frac{1}{2}}.
\end{eqnarray}
From Eqs.(\ref{12}) and (\ref{25}), we obtain the kinetic energy
term as
\begin{eqnarray}\nonumber
\phi'(a)&=&\frac{1}{H_{0}}\left[\frac{2(\mu-\lambda)\Omega_{k_{0}}}{3(\mu-\lambda-1)}e^{-2x}
-\frac{(2\mu-3(1-d^2)\lambda-2d^2)\Omega_{m_{0}}}{2(\mu-1)-3(1-d^2)\lambda}e^{-3(1
-d^2)x}\right.\\\nonumber
&+&\left.\frac{2b(\mu-1)}{3\lambda}e^{\frac{-2(\mu-1)x}{\lambda}}\right]^{\frac{1}{2}}
\left[\frac{(\mu-\lambda)\Omega_{k_{0}}}{(\mu-\lambda-1)}e^{-2x}-(2\mu-3(1-d^2)\lambda)\right.\\\nonumber
&\times&\left.(2(\mu-1)-3(1-d^2)\lambda)^{-1}\Omega_{m_{0}}e^{-3(1-d^2)x}
+be^{\frac{-2(\mu-1)x}{\lambda}}\right]^{-\frac{1}{2}}\left[\Omega_{k_{0}}\right.\\\label{27}
&\times&\left.(\mu-\lambda-1)^{-1}e^{-2x}
-\frac{2\Omega_{m_{0}}e^{-3(1-d^2)x}}{2(\mu-1)-3(1-d^2)\lambda}+be^{\frac{-2(\mu-1)x}{\lambda}}\right]^{-\frac{1}{2}}.
\end{eqnarray}

The plot of the tachyon field $\phi$ versus $x$ is given in Figure
\textbf{6} which shows that the scalar field $\phi$ increases with
the passage of time and becomes steeper for increasing the
interacting parameter $d^2$. However, the kinetic energy of the
tachyon potential decreases and approaches to zero in the future.
With this behavior, Eq.(\ref{24}) indicates the vacuum evolution of
the universe. The tachyon potential shows oscillation initially
about its maxima (it increases with the increment of $d^2$) but
approaches to zero in the later time as shown in Figure \textbf{7}.
For the later time, its rapid decrease from maxima gives inverse
proportionality to the scalar field. This type of behavior
corresponds to scaling solutions in the brane-world cosmology
\cite{30}.

\subsection{New Holographic K-essence Model}

\begin{figure} \centering
\epsfig{file=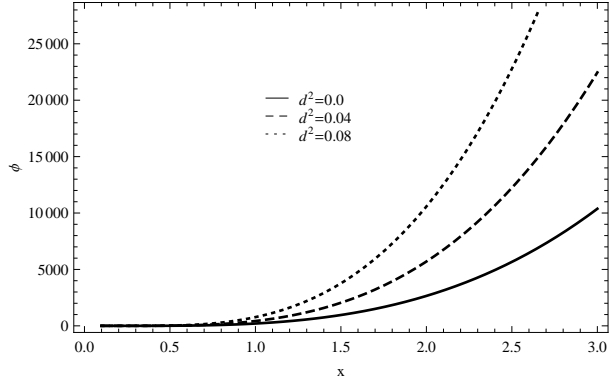,width=.60\linewidth} \caption{Plot of $\phi$
versus $x$ for tachyon model.}
\end{figure}
\begin{figure} \centering
\epsfig{file=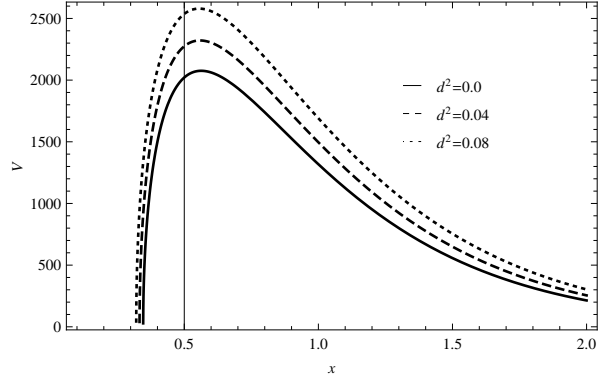,width=.60\linewidth} \caption{Plot of $V$ versus
$x$ for tachyon model.}
\end{figure}
The K-essence model is different from quintessence scalar field
model in the sense that it evolutes the universe in the accelerated
expansion era. It is originated from the idea of K-inflation which
was used to describe the inflation of the early universe at high
energies \cite{21}. This model has been used as an alternative
candidate of DE which yields interesting results of scaling and
attractor solutions \cite{32,33}. This model is described with a
scalar field having non-canonical kinetic energy. The generalized
form of scalar field action is \cite{22}
\begin{equation*}
S=\int{d^4x\sqrt{-g}}p(\phi,\chi),
\end{equation*}
where $p(\phi,\chi)$ shows the pressure density as a function of
potential $\phi$ and $\chi=\frac{1}{2}~\dot{\phi}^2$. The
corresponding energy density and pressure are
\begin{eqnarray}\label{28}
\rho_{k}=V(\phi)(-\chi+3\chi^{2}),\quad
p_{k}=V(\phi)(-\chi+\chi^{2}),
\end{eqnarray}
where $V(\phi)$ represents the scalar potential of K-essence model.
The corresponding EoS parameter is
\begin{equation}\label{30}
\omega_{k}=\frac{1-\chi}{1-3\chi}.
\end{equation}
Equating $\rho_{k}=\rho_{\Lambda}$ and $p_{k}=p_{\Lambda}$ for the
correspondence between NHDE and K-essence model, we obtain
\begin{eqnarray}\nonumber
\chi&=&\left[\frac{4(\mu-\lambda)\Omega_{k_{0}}}{3(\mu-\lambda-1)}e^{-2x}
-\frac{(2\mu-3(1-d^2)\lambda+2d^2)\Omega_{m_{0}}}{2(\mu-1)-3(1-d^2)\lambda}e^{-3(1
-d^2)x}\right.\\\nonumber
&+&\frac{b}{3\lambda}\left.(6\lambda-2\mu+2)e^{\frac{-2(\mu-1)x}{\lambda}}\right]
\left[\frac{2(\mu-\lambda)\Omega_{k_{0}}}{(\mu-\lambda-1)}e^{-2x}
-(2\mu-3(1-d^2)\lambda\right.\\\nonumber
&+&\left.6d^2)(2(\mu-1)-3(1-d^2)\lambda)^{-1}\Omega_{m_{0}}e^{-3(1-d^2)x}
+\frac{b}{\lambda}(4\lambda-2\mu+2)\right.\\\label{31}
&\times&\left.e^{\frac{-2(\mu-1)x}{\lambda}}\right]^{-1},\\\nonumber
V(\phi)&=&\frac{3(1-3\omega_{k})^2H^{2}_{0}}{2(1
-\omega_{k})}\left[\frac{(\mu-\lambda)\Omega_{k_{0}}}{\mu-\lambda-1}e^{-2x}
-\frac{(2\mu-3(1-d^2)\lambda)\Omega_{m_{0}}}{2(\mu-1)-3(1-d^2)\lambda}\right.\\\label{32}
&\times&\left.e^{-3(1
-d^2)x}+be^{\frac{-2(\mu-1)x}{\lambda}}\right].
\end{eqnarray}
\begin{figure} \centering
\epsfig{file=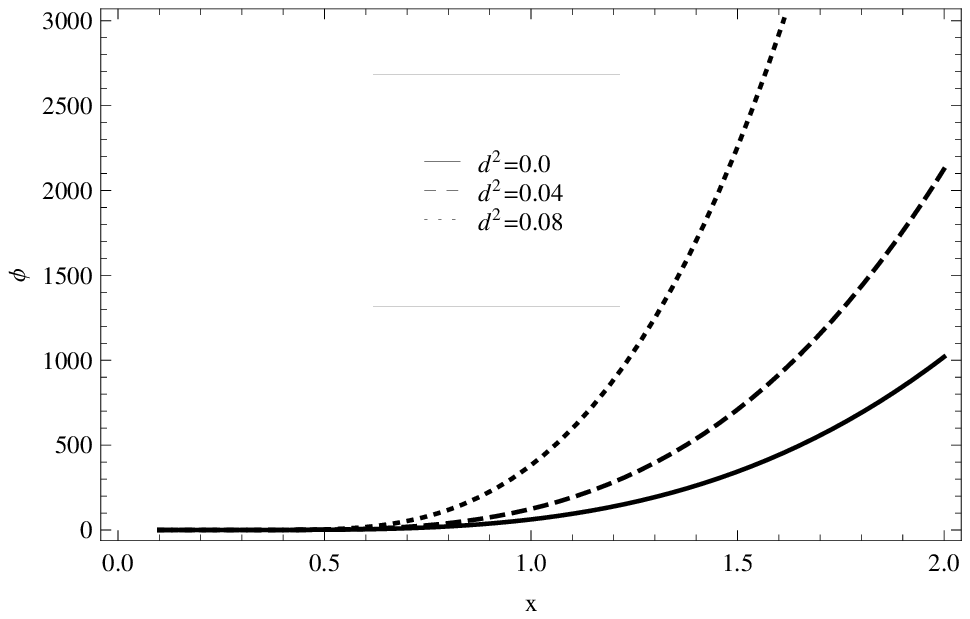,width=.60\linewidth} \caption{Plot of $\phi$
versus $x$ for K-essence model.}
\end{figure}
\begin{figure} \centering
\epsfig{file=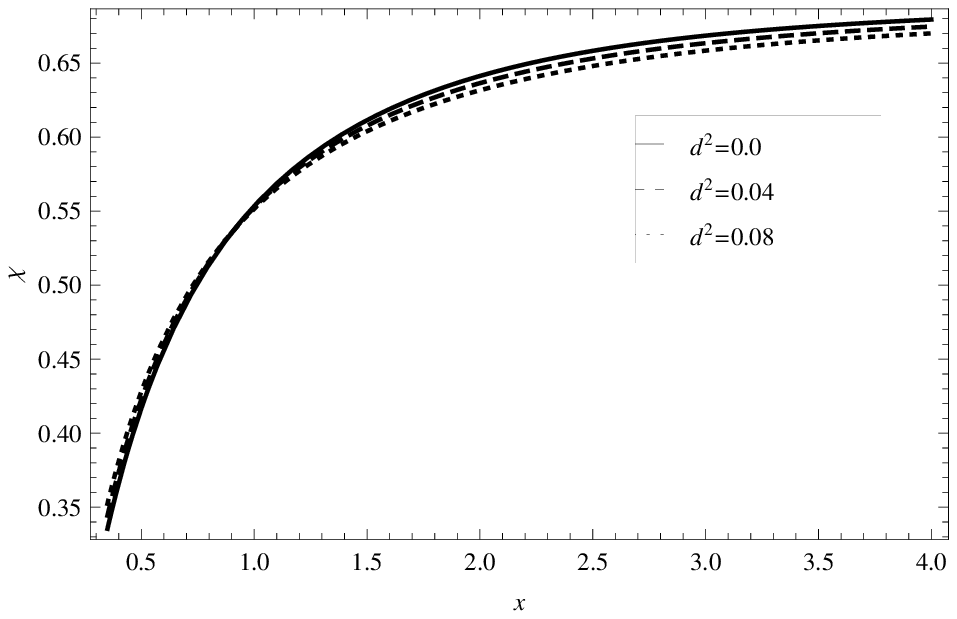,width=.60\linewidth} \caption{Plot of $\chi$
versus $x$ for K-essence model.}
\end{figure}
\begin{figure} \centering
\epsfig{file=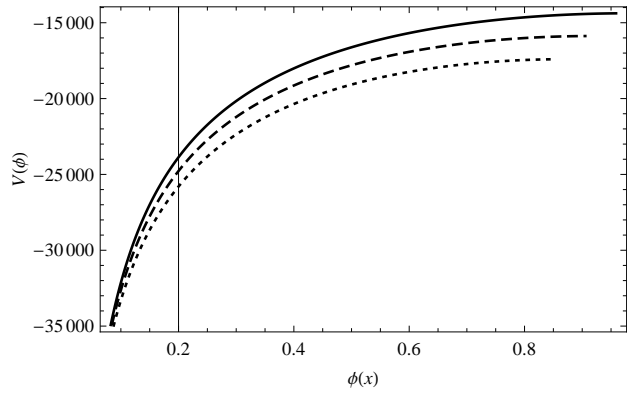,width=.60\linewidth} \caption{Plot of $V$ versus
$\phi$ for K-essence model.}
\end{figure}

Consequently, the relation $\chi=\frac{1}{2}\dot{\phi}^2$ gives the
evolution
\begin{eqnarray}\nonumber
\phi'(a)&=&\frac{\sqrt{2}}{H_{0}}\left[\frac{4(\mu-\lambda)\Omega_{k_{0}}}{3(\mu-\lambda-1)}e^{-2x}
-\frac{(2\mu-3(1-d^2)\lambda+2d^2)\Omega_{m_{0}}}{2(\mu-1)-3(1-d^2)\lambda}e^{-3(1
-d^2)x}\right.\\\nonumber
&+&\left.\frac{b(6\lambda-2\mu+2)}{3\lambda}e^{\frac{-2(\mu-1)x}{\lambda}}\right]^\frac{1}{2}
\left[\frac{2(\mu-\lambda)\Omega_{k_{0}}}{(\mu-\lambda-1)}e^{-2x}-(2\mu-3(1-d^2)\lambda\right.\\\nonumber
&+&\left.6d^2)(2(\mu-1)-3(1-d^2)\lambda)^{-1}
\Omega_{m_{0}}e^{-3(1-d^2)x}+\frac{b}{\lambda}(4\lambda-2\mu+2)\right.\\\nonumber
&\times&\left.e^{\frac{-2(\mu-1)x}{\lambda}}\right]^{-\frac{1}{2}}\left[\frac{\Omega_{k_{0}}}
{(\mu-\lambda-1)}e^{-2x}-2\Omega_{m_{0}}e^{-3(1-d^2)x}(2(\mu-1)\right.\\\label{33}
&-&\left.3(1-d^2)\lambda)^{-1}+be^{\frac{-2(\mu-1)x}{\lambda}}\right]^{-\frac{1}{2}}.
\end{eqnarray}
From here we can plot $\phi$ and $\chi$ versus $x$ with previous
assumptions. We see from Figure \textbf{8} that the scalar field
$\phi$ increases. Figure \textbf{9} shows that $\chi$ almost lies in
the required interval $\left(\frac{1}{3},\frac{2}{3}\right)$ from
early epoch to the later time. The EoS parameter (\ref{30})
indicates that the accelerated universe can be obtained for this
interval. For $\chi<\frac{1}{2}$, it gives phantom DE era which
corresponds to late time attractor \cite{32}. Figure \textbf{10}
shows that $V(\phi)$ increases slowly but attains very large
negative value with the increase of scalar field $\phi$.

\subsection{New Holographic Dilaton Field}

The Lagrangian of dilaton field can be expressed in terms of
pressure of scalar field as \cite{21}
\begin{equation}\label{34}
p_d=-\chi+b_{1}e^{b_2\phi}\chi^2,
\end{equation}
where $b_{1}$ and $b_{2}$ are taken as positive constants. This
Lagrangian (pressure) produces the following energy density
\begin{eqnarray}\label{35}
\rho_{d}&=&-\chi+3b_{1}e^{b_{2}\phi}\chi^{2}.
\end{eqnarray}
The EoS for dilaton DE is
\begin{equation}\label{36}
\omega_{d}=\frac{-1+b_{1}e^{b_{2}\phi}\chi}{-1+3b_{1}~e^{b_{2}\phi}\chi}.
\end{equation}
By setting $\rho_{d}=\rho_{\Lambda}$ and $p_{d}=p_{\Lambda}$, we
obtain
\begin{eqnarray}\nonumber
e^{b_{2}\phi}\chi&=&\frac{1}{b_{1}}\left[\frac{4(\mu-\lambda)\Omega_{k_{0}}}{3(\mu-\lambda-1)}e^{-2x}
-\frac{(2\mu-3(1-d^2)\lambda+2d^2)\Omega_{m_{0}}}{2(\mu-1)-3(1-d^2\lambda)}e^{-3(1
-d^2)x}\right.\\\nonumber
&+&\left.\frac{b(6\lambda-2\mu+2)}{3\lambda}e^{\frac{-2(\mu-1)x}\lambda}\right]
\left[\frac{2(\mu-\lambda)\Omega_{k_{0}}}{\mu-\lambda-1}~e^{-2x}
-\Omega_{m_{0}}e^{-3(1-d^2)x}\right.\\\label{37}
&\times&\left.\frac{(2\mu-3(1-d^2)\lambda+6d^2)}{2(\mu-1)-3(1-d^2)\lambda}
+\frac{b}{\lambda}(4\lambda-2\mu+2)e^{\frac{-2(\mu-1)x}{\lambda}}\right]^{-1}.
\end{eqnarray}

The EoS parameter (\ref{36}) gives the bound of $e^{b_{2}\phi}\chi$
which is $\left(\frac{20}{3},\frac{40}{3}\right)$ in order to obtain
the accelerated universe. Its graph versus $x$ with $\alpha=0.05$ is
shown in Figure \textbf{11} which indicates that this almost lies in
the same interval and shows consistency. Also, the solution of
(\ref{37}) follows
\begin{figure} \centering
\epsfig{file=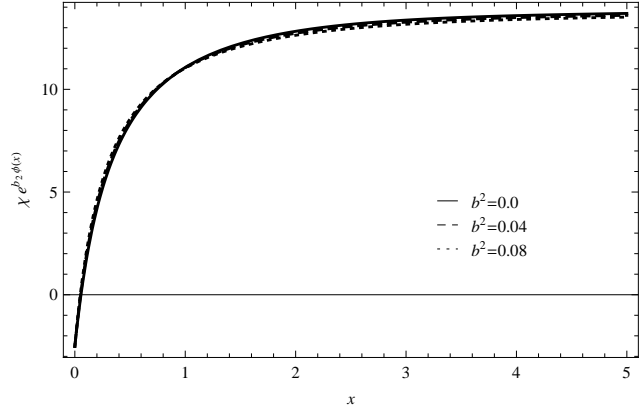,width=.65\linewidth} \caption{Plot of
$e^{b_{2}\phi}\chi$ versus $x$ for dilaton field.}
\end{figure}
\begin{eqnarray}\nonumber
\phi(a)&=&\frac{2}{b_{2}}\ln\left[1+\frac{b_{2}}{\sqrt{2b_{1}H^2_{0}}}\int^{x}_{0}
\left[\frac{4(\mu-\lambda)\Omega_{k_{0}}}{3(\mu-\lambda-1)}e^{-2x}
-\Omega_{m_{0}}e^{-3(1 -d^2)x}\right.\right.\\\nonumber
&\times&\left.\left.\frac{(2\mu-3(1-d^2)\lambda+2d^2)}{2(2(\mu-1)-3(1-d^2)\lambda)}
+\frac{b(6\lambda-2\mu+2)}{3\lambda}
e^{\frac{-2(\mu-1)x}{\lambda}}\right]^\frac{1}{2}\right.\\\nonumber
&\times&\left.\left[\frac{2(\mu-\lambda)\Omega_{k_{0}}}{(\mu-\lambda-1)}e^{-2x}
-\frac{(2\mu-3(1-d^2)\lambda+6d^2)\Omega_{m_{0}}}{2(\mu-1)-3(1-d^2)\lambda}
e^{-3(1-d^2)x}\right.\right.\\\nonumber &+&\left.\left.
\frac{b}{\lambda}(4\lambda-2\mu+2)be^{\frac{-2(\mu-1)x}{\lambda}}\right]^{-\frac{1}{2}}
\left[\frac{\Omega_{k_{0}}}{(\mu-\lambda-1)}e^{-2x}-2\Omega_{m_{0}}
\right.\right.\\\label{38}
&\times&\left.\left.(2(\mu-1)-3(1-d^2)\lambda)^{-1}e^{-3(1-d^2)x}
+be^{\frac{-2(\mu-1)x}{\lambda}}\right]^{-\frac{1}{2}}dx\right].
\end{eqnarray}
\begin{figure} \centering
\epsfig{file=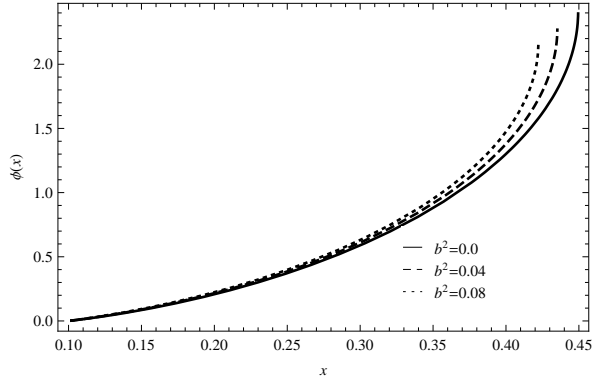,width=.60\linewidth} \caption{Plot of $\phi$
versus $x$ for dilaton field.}
\end{figure}
Its graph is shown in Figure \textbf{12} which exhibits direct
proportionality with respect to $x$. This type of behavior gives
scaling solutions for dilaton model as proved in \cite{21}.

\section{Generalized Second Law of Thermodynamics}

In this section, we check the validity of GSLT for interacting NHDE
in the non-flat universe. Bekenstein's \cite{34} provided a
relationship about the entropy of black hole horizon and horizon
area which plays a crucial role to discuss the GSLT. This law states
that the sum of black hole entropy and the background entropy must
increase with the passage of time. The first law of thermodynamics
yields
\begin{equation}\label{39}
TdS=pdV+dE,
\end{equation}
where $T,~S,~p,~V$ and $E$ denote temperature, entropy, pressure,
volume and internal energy of the system, respectively.
Differentiating it with respect to time, we obtain
\begin{equation}\label{40}
\dot{S_{\Lambda}}=\frac{p_{\Lambda}\dot{V}+\dot{E_{\Lambda}}}{T},\quad
\dot{S_{m}}=\frac{p_{m}\dot{V}+\dot{E_{m}}}{T},
\end{equation}
for the NHDE and DM, respectively. The volume, temperature and
entropy of horizon $L$ in non-flat universe are defined as \cite{35}
\begin{equation}\label{41}
V=\frac{4\pi~L^{3}}{3},\quad T=\frac{1}{2\pi L},\quad
S_{H}=2\pi^{2}L^{2}.
\end{equation}
The internal energies for NHDE and DM are
\begin{eqnarray}\label{42}
E_{\Lambda}=\frac{4\pi L^{3}\rho_{\Lambda}}{3},\quad
E_{m}=\frac{4\pi L^{3}\rho_{m}}{3}.
\end{eqnarray}
Using Eqs.(\ref{8}), (\ref{10}) and (\ref{40})-(\ref{42}), we have
the final expression of the GSLT
\begin{eqnarray}\nonumber
S'_{total}&=&-\frac{4\pi^{2}}{9\rho^3_{\Lambda}}[(1+\omega_{\Lambda})\rho_{\Lambda}
+3H^2_{0}\Omega_{m_0}e^{-3(1-d^2)x}](\rho'_{\Lambda}+2\rho_{\Lambda})
-\frac{\pi^{2}\rho'_{\Lambda}}{3\rho_{\Lambda}^{2}},
\end{eqnarray}
where $S'_{total}$ denotes the derivative of the sum of entropies
due to DM, NHDE and horizon entropy. We plot it against $x$ with the
same assumptions as given earlier. Figure \textbf{13} shows that the
GSLT is valid for early epoch but fails for the later time in this
scenario.
\begin{figure} \centering
\epsfig{file=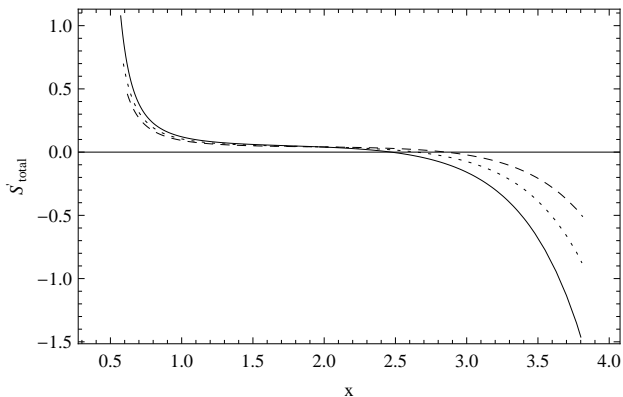,width=.60\linewidth} \caption{Plot of
$S'_{total}$ versus $x$ for NHDE.}
\end{figure}

\section{Concluding Remarks}

It was shown through experimental data that our universe is not
totally flat but it contains small positive curvature
\cite{25a}-\cite{25i}. This motivated us to perform the versatile
study of the interacting NHDE in the context of non-flat universe.
In this paper, we have discussed four main features by choosing
different values of interacting parameter i.e., $d^2=0, 0.04, 0.08$
which are summarized as follows.  Firstly, we obtain the evolution
equation which starts from vacuum DE region $(\omega_{\Lambda}=-1)$
and goes towards quintessence region $(\omega_{\Lambda}>-1)$ as
shown in Figure \textbf{1}. This type of behavior is consistent with
the present observations. Secondly, we have discussed the
instability of this model against perturbation as shown in Figure
\textbf{2}. It is found that it remains unstable forever like HDE
with IR cutoff as a future event horizon \cite{18}, agegraphic
\cite{19} and ghost QCD \cite{20} DE models but unlike Chaplygin gas
\cite{17} DE model.

Thirdly, we have evaluated the interacting NHDE versions of
quintessence, tachyon, K-essence and dilaton scalar field DE models.
We have also explored the behavior of potentials and dynamics of the
scalar field corresponding to these models by using the reliable
values of the constant parameters. These are shown in Figures
\textbf{3-12}. It is seen that the scalar field $\phi$ becomes more
steeper with the increase of the interacting parameter $d^2$ for all
scalar field models. Also, the scalar potential $V(\phi)$ decreases
more rapidly in case of quintessence for increasing the interacting
parameter $d^{2}$ as compared to non-interacting scenario (Figure
\textbf{{5}}). However, for K-essence model, it increases but
achieves a maximum value lower than that of non-interacting case as
shown in Figure \textbf{{10}}. It was argued \cite{22,36,37} that
the scalar field models favor the DE phenomenon for EoS parameter
lying in the interval $(-1,0)$. However, in our case, the EoS of
NHDE remains in this region and shows consistency.

We would like to mention here that our solutions coincide with the
attractor solutions (for quintessence \cite{22} and K-essence
\cite{32} DE models) and scaling solutions (for tachyon \cite{30}
and dilaton \cite{21} DE models). It is remarked that our
expressions of EoS parameter and scalar field models can be reduced
to the results of \cite{23} (with vanishing of DM density and $k=0$)
and of \cite{24} (for the vanishing of DM density only). Finally, we
have checked the validity of the GSLT in this scenario. It is found
that initially it is valid but fails for the later time.


\begin{thebibliography}{43}

\bibitem{1} Perlmutter, S. et al.: Astrophys. J. \textbf{517}(1999)565.

\bibitem{2} Caldwell, R.R. and Doran, M.: Phys. Rev. D \textbf{69}(2004)103517.

\bibitem{3} Koivisto, T. and Mota, D.F.: Phys. Rev. D \textbf{73}(2006)083502

\bibitem{4} Daniel, S.F.: Phys. Rev. D \textbf{77}(2008)103513.

\bibitem{5} Hoekstra, H. and Jain, B.: Ann. Rev. Nucl. Part. Sci. \textbf{58}(2008)99.

\bibitem{6} Fedeli, C., Moscardini, L. and Bartelmann, M.: Astron. Astrophys.
\textbf{500}(2009)667.

\bibitem{7} Peebles, P.J.E.: Rev. Mod. Phys. \textbf{75}(2003)559.

\bibitem{8} Susskind, L.: J. Math. Phys. \textbf{36}(1995)6377.

\bibitem{9} Cohen, A., Kaplan, D. and Nelson, A.: Phys. Rev. Lett. \textbf{82}(1999)4971.

\bibitem{10} Hsu, S.D.H.: Phys. Lett. B \textbf{594}(2004)13.

\bibitem{11} Li, M.: Phys. Lett. B \textbf{603}(2004)1.

\bibitem{12} Gao, C., Chen, X. and Shen, Y.G.: Phys. Rev. D \textbf{79}(2009)043511.

\bibitem{13} Granda, L. and Oliveros, A.: Phys. Lett. B \textbf{669}(2008)275.

\bibitem{14} Chen, S. and Jing, J.: Phys. Lett. B \textbf{679}(2009)144.

\bibitem{15} Zhang, X.: Phys. Rev. D \textbf{79}(2009)103509;
Wang, Y. and Xu, L.: Phys. Rev. D \textbf{81}(2010)083523.

\bibitem{17} Gorini, V. et al.: Phys. Rev. D \textbf{72}(2005)103518;
Sandvik, H. et al.: Phys. Rev. D \textbf{69}(2004)123524.

\bibitem{18} Myung, Y.S.: Phys. Lett. B \textbf{652}(2007)223.

\bibitem{19} Kim, K.Y., Lee, H.W. and Myung, Y.S.: Phys. Lett. B \textbf{660}(2008)118.

\bibitem{20} Ebrahimi, E. and Sheykhi, A.: Int. J. Mod. Phys. D
\textbf{20}(2011)2369.

\bibitem{21} Piazza, F. and Tsujikawa, S.: JCAP \textbf{07}(2004)004.

\bibitem{22} Copeland, E.J., Sami, M. and Tsujikawa, S.: Int. J. Mod.
Phys. D \textbf{15}(2006)1753.

\bibitem{23} Granda, L. and Oliveros, A.: Phys. Lett. B \textbf{671}(2009)199.

\bibitem{24} Karami, K. and Fehri, J.: Phys. Lett. B \textbf{684}(2010)61.

\bibitem{27} Sheykhi, A.: Phys. Rev. D \textbf{84}(2011)107302.

\bibitem{25} Setare, M.R.: JCAP \textbf{01}(2007)023;
Sheykhi, A.: Class. Quantum Grav. \textbf{27}(2010)025007; Mazumder,
M. and Chakraborty, S.: Gen. Relativ. Gravit. \textbf{42}(2010)813.

\bibitem{25a} Huang, Q.G. and Li, M.: JCAP \textbf{08}(2004)013.

\bibitem{25b} Bennett, C.L. et al.: Astrophys. J. Suppl.
\textbf{148}(2003)1.

\bibitem{25c} Sievers, J.L. et al.: Astrophys. J. \textbf{591}(2003)599.

\bibitem{25d} Efstathiou, G.: Mon. Not. Roy. Astron. Soc. \textbf{343}(2003)L95.

\bibitem{25e} Luminet, J.P.: Nature \textbf{425}(2003)593.

\bibitem{25f} Tegmark, M. et al.: Phys. Rev. D \textbf{69}(2004)103501.

\bibitem{25g} Gong, Y., Wang, B. and Zhang, Y.Z.: Phys. Rev. D \textbf{72}(2005)043510.

\bibitem{25h} Seljak, U., Slosar, A. and McDonald, P.: JCAP \textbf{10}(2006)014.

\bibitem{25i} Spergel, D.N. et al.: Astrophys. J. Suppl. \textbf{170}(2007)377.

\bibitem{25j} Ichikawa, K. et al.: JCAP \textbf{06}(2006)005.

\bibitem{25k} Lu, J. et al.: JCAP \textbf{03}(2010)031.

\bibitem{28} Mazumdar, A., Panda, S. and Perez-Lorenzana, A.: Nucl.
Phys. B \textbf{614}(2001)101; Gibbons, G.W.: Phys. Lett. B
\textbf{537}(2002)1; Feinstein, A.: Phys. Rev. D
\textbf{66}(2002)063511; Piao, Y.S. et al.: Phys. Rev. D
\textbf{66}(2002)121301.

\bibitem{30} Tsujikawa, S. and Sami, M.: Phys. Lett. B
\textbf{603}(2004)113; Mizuno, S., Lee, S.J. and Copeland, E.J.:
Phys. Rev. D \textbf{70}(2004) 043525.

\bibitem{32} Chiba, T., Okabe, T. and Yamaguchi, M.: Phys. Rev. D \textbf{62}(2000)023511.

\bibitem{33} Armend$\acute{a}$riz-Pic$\acute{o}$n, C., Mukhanov, V. and Steinhardt, P.J.:  Phys.
Rev. Lett. \textbf{85}(2000)4438; Phys. Rev. D
\textbf{63}(2001)103510.

\bibitem{34} Bekenstein, J.D.: Phys. Rev. D \textbf{7}(1973)2333.

\bibitem{35} Cai, R.G. and Kim, S.P.: JHEP \textbf{0502}(2005)050.

\bibitem{36} Zhang, X.: Phys. Lett. B \textbf{648}(2007)1; Phys.
Rev. D \textbf{74}(2006)103505.

\bibitem{37} Rozas-Fern$\acute{a}$ndez, A.: Eur. Phys. J. C \textbf{71}(2011)1536.
\end{thebibliography}
\end{document}